\documentclass[letterpaper, 10 pt, conference]{ieeeconf}  
\usepackage{FG2023}
\usepackage{caption}
\usepackage{subcaption}
\usepackage{graphicx}
\usepackage[noadjust,sort]{cite}

\FGfinalcopy 

\def\FGPaperID{16} 

\title{\LARGE \bf
The Role of Vocal Persona in Natural and Synthesized Speech}

\author{\parbox{16cm}{\centering
    {\large Camille Noufi, Lloyd May and Jonathan Berger}\\
    {\normalsize
    Center for Computer Research in Music and Acoustics, Stanford University, Stanford, CA, USA\\}}
}

\begin{document}

\ifFGfinal
\thispagestyle{empty}
\pagestyle{empty}
\else
\author{Anonymous FG2023 submission\\ Paper ID \FGPaperID \\}
\pagestyle{plain}
\fi
\maketitle

\begin{abstract}
The inclusion of voice persona in synthesized voice can be significant in a broad range of human-computer-interaction (HCI) applications, including augmentative and assistive communication (AAC), artistic performance, and design of virtual agents. 
We propose a framework to imbue compelling and contextually-dependent expression within a synthesized voice by introducing the role of the vocal persona within a synthesis system.  
In this framework, the resultant ‘tone of voice’ is defined as a point existing within a continuous, contextually-dependent probability space that is traversable by the user of the voice. 
We also present initial findings of a thematic analysis of 10 interviews with vocal studies and performance experts to further understand the role of the vocal persona within a natural communication ecology.
The themes identified are then used to inform the design of the aforementioned framework.

\end{abstract}


\section{Introduction}\label{sec:intro}

The human ability to control and manipulate one's voice both linguistically and paralinguistically facilitates vocal communication of thoughts and feelings, embodying individual personality traits with subtlety~\cite{Scherer1978PersonalityExtroversion, Apple1979, Campbell2005GettingLanguage,Kreiman2011FoundationsPerception}, enhancing the voice's central role in human social interaction.
Adaptive modulation of the voice to convey expression and communicate with contextually-appropriate specificity is a critical aspect of human interaction.
Our purpose here is to clarify the relationship between environmental context and vocal \textit{expressivity} both behaviorally and within a mathematical framework, with the aim of improving context-aware expressivity in voice synthesis systems by characterizing and describing \textit{vocal persona}.

The past several decades of speech synthesis research have seen great progress in modeling and implementing paralinguistic control of the voice~\cite{Serra1990,Yoshimura1999,Zen2009,Wang2018,Hsu2019,Valle2020Flowtron:Synthesis,Valle2020mellotron}. 
Additionally, recent advances in expressive neural TTS research have enabled control of low-level prosodic features (e.g. speech rate, voice quality, and pitch range) in addition to selection of speaker identity and speaking style.
Vocal attributes manipulated for expressivity such as pitch variability, voice quality, pronunciation/stress, and speech rate/cadence can be determined by example~\cite{Wang2018, Valle2020Flowtron:Synthesis, Shechtman2021SynthesisArchitecture} or by direct control~\cite{Sorin2017SemiCapabilities,Wang2018,Hsu2019,Valle2020mellotron, Morrison2021Context-AwareEditing, Morrison2021NeuralLPCNet,Neekhara2021ExpressiveCloning}.
Furthermore, the role of personality and expressivity of synthetic voice in social interaction has previously been shown to increase technologically-mediated engagement and connection~\cite{Pullin201517Technology, Higginbotham2010HumanizingCommunication, Fiannaca2018Voicesetting:Communication, Stan2021GeneratingAssistant, Zhang2021SocialVoice}.  
However, several key aspects of paralinguistic control, such as contextual awareness and user-determined parameter management, demand further study in speech synthesis research.
Furthermore, a more nuanced approach is needed to enable speech synthesis to meaningfully explore the relationship between speaker identity and adjustable prosodic control.

The notion of \textit{vocal persona} is described by Tagg as the capacity to control the voice ``not just to utter words but also to present our individual or group identity, and to express emotions, attitudes and behavioural positions"~\cite{Tagg2012VocalPersona}. It is a vocal manifestation of \textit{persona}---defined as the aspect of someone's character that is presented to or perceived by others---made perceptible via vocal prosody or singing.
In this short paper, we present our initial design framework for expressive vocal synthesis that leverages a voice persona, drawn from an underlying continuous probability space, that bounds and contextualizes low-level/latent synthesis features. The framework enables user-defined perceptual abstraction that could allow for real-time expressive manipulation within and between chosen voice personas. 

By first gaining a deeper understanding of how a vocal persona functions within a natural communication ecology, we can further specify its role within a synthetic speech system.
We are unaware of prior work on the role of vocal persona within communities of voice teachers, or professional speakers and singers. 
In addition, to our knowledge, there is no scholarly work on vocal persona in the speech technologies communities.
Despite this absence, the notion of ``vocal persona" is often discussed by singers, actors, conversational voice designers, and voiceover artists.
Thus, we conduct semi-structured interviews with ten professional vocal artists to gather such insights from the voice community. 
Through these interviews, we seek to understand (1) how the expressive voice functions within one's relationship to their environment, and (2) how artists describe physical and cognitive vocal embodiment, sensory feedback, and vocal manipulation. 
We present initial resulting themes that influence and shape vocal persona within natural communication and identify three organizing principles of vocal persona that bridge environmental context with vocal expression.
To conclude the paper, we discuss broader insights gained into how HCI elements of expressive synthesized voice control may continued to be improved.


\section{Thematic Analysis of Persona \\Within Natural Voice}\label{sec:thematic-analysis}

For this pilot study, interview participants were recruited via email from a community of performers and professionals studying the human voice. 
Qualitative data was collected through semi-structured interviews conducted online through Zoom video conferencing software, offering flexibility of topic-focus to participants.
We use thematic analysis as an analysis framework of expert interview data, as it specifically allows for the inclusion of an initial hypothesis through a definition of a priori themes that are then iteratively modified throughout analysis~\cite{terry2017thematic}.
Participants were asked questions pertaining to a priori topics of interest, namely: physical context, sociocultural context~\cite{Kreiman2011FoundationsPerception, Sweet2019FemalePhenomenology}, technological context~\cite{Park2020BeyondOnline}, perception of the self and other~\cite{Kreiman2011FoundationsPerception, Sweet2019FemalePhenomenology}, and perception of agency~\cite{Linklater2006FreeingLanguage}.

\subsection{Resulting Themes}\label{sec:thematic-results}
It was clear throughout the ten interviews that vocal persona plays a key role in both participants' relationship to human voice and vocal modifications in response to various internal and external factors.
All participants mentioned between 2 and 7 (mean = 4.8) different vocal personas they would adopt in response to various common social and physical contexts. 
Additionally, they mentioned adjusting their vocal persona in response to a social relationship between 2 and 5 times (mean = 3.3) during the interview.

Iterative thematic analysis of participant interviews yielded six concepts as having a strong influence on the usage of a vocal persona in an expressive communication setting.
\textit{\textbf{Physical context:}}  
Real-time self-adjustment would often happen in response to acoustic effects of the physical space, background noise, and/or physical proximity to others. 
This self-adjustment often informed the adoption of a more pointedly articulated or intelligible vocal tone within a vocal persona.
\textit{\textbf{Technological Mediation}} seems to influence vocal production choices.  
Inclusion or exclusion of the audiovisual or haptic modalities impacted the salience of paralinguistic cues included in vocal tone to convey emotion, intelligibility or intent.  
Awareness of and proximity to a vocal amplifier such as a microphone greatly influenced the inclusion of specific vocal personas and unique modes of speech.
\textit{\textbf{Voice Acoustics}} were discussed directly, metaphorically, and with physiological terms.
As the participants all had 10+ years of vocal experience, they were able to describe modification of their voice using haptic, acoustic and anatomical terminology, but often also relied on simile and metaphor to discuss expressive vocal tones. 
It was very clear that physicality of the vocal signal within the body was an extremely important feedback mechanism to inform self-awareness of and personal relationship to the voice, especially in new or unfamiliar settings.
\textit{\textbf{Self-Perception}} questions uncovered a prevalent concept of a set of vocal tones and production modes that together yielded a baseline vocal identity. This baseline vocal identity was often referred to as “my authentic voice.”  
Participants were aware that their baseline vocal identity was a smaller subset of all possible vocalizations and vocal modes a person could physically produce.
Ubiquitously, there was a disconnect between internal and external (e.g. via a recording or descriptions from others) auditory and haptic perceptions of one's vocal tone. Exposure over time to this disconnect increased comfort with externalized self-perception.
\textit{\textbf{Perception of Others:}} All participants said they make assumptions about another person’s physicality, personality, intelligence, or experiences based on their voice.  
Additionally, assumptions of a vocal tone were often made based on physical attributes of a person.  
These were often informed by social scripting, entrainment and expectations.
\textit{\textbf{Sociocultural Context}} and \textit{\textbf{Performativity}}: 
Social structures such as behavioral and power dynamic roles impacted vocal production.  
This informed adoption of characterological vocal patterns to uphold such roles~\cite{eckert2019limits}.
Internal emotional states were often modulated or masked by the speaker, depending on the sociocultural contexts.
In line with the findings of Kreiman~\cite{Kreiman2011FoundationsPerception}, it was apparent that there can often be a conscious or subconscious disconnect between felt emotions and those expressed through the voice.
Describing the relationship between vocal emotion and personality, Kreiman argues that, while humans share many vocal emotion processes with animals, it is our elaborated cognitive processes that allow us to consciously modulate what is produced.
This ability to consciously modulate our voices allows us to control what is perceived by others.
Performativity was employed as a mean to manipulate vocal tone and cadence when communicating an intention or expressing an internal state, as well as when creating or adapting an identity. 
Adoptions of these vocal patterns highlighted the continuous acoustic space between the ``typical voice," inside which many personas were employed, as well as that of the voice used to embody a character or archetype. 
Vocal performers had a heightened sense of awareness and duty around how the manipulated or managed voice would be perceived and exactly what to modify in the voice and body to accomplish this.
  
We found three main organizing principles influencing vocal persona present across the themes above, namely: information hierarchy, type of mediation, and agency.  
Decisions around \textit{information hierarchy} influenced vocal persona by determining prioritization of communicated semantic information, emotional or internal state, and shared referential concepts (such as subtle vocal references to a characterological figure or archetype).
The \textit{method and degree of vocal mediation} influenced vocal persona by bounding certain vocal production choices deemed necessary to align with their surrounding physical, sociocultural and/or technological context.  
\textit{Agency} was an extremely common concept discussed around the decision to adopt a particular vocal persona. 
Information hierarchy and contextual mediation provided cues as to optimal expression of information or internal state through the voice, but the adoption and employment of the vocal persona were heavily impacted by the speaker's awareness of the possibilities of vocal manipulation both within and external to their ``typical" self-perceived voice. 
The agency of a speaker to consciously adjust to their surrounding context was both important and ubiquitous.

\section{Persona-Informed Synthesis Framework}\label{sec:framework}

Drawing from theories of performativity~\cite{butler1990gender}, vocal code-switching~\cite{Bullock2009}, and the themes presented in Section \ref{sec:thematic-results}, we propose that a vocal persona is sampled from a fluid \textit{persona probability space} that contextualizes the voice one may use in a certain setting or to embody a certain personality.\footnote{We also reference digital musical instrument (DMI) design~\cite{cook2004}, which provides user agency over acoustic nuance within and between timbral spaces.}
For example, one vocal persona adopted for ``meeting with clients" and another adopted for ``chatting with family" perhaps share a similar but differing distribution space.
In contrast, a user's chosen voice for ``chatting with family" may have much less overlap with a ``delivering a speech" persona.

\begin{figure}[tb!]
     \centering
     \centerline{\framebox{
     \includegraphics[width=.7\columnwidth]{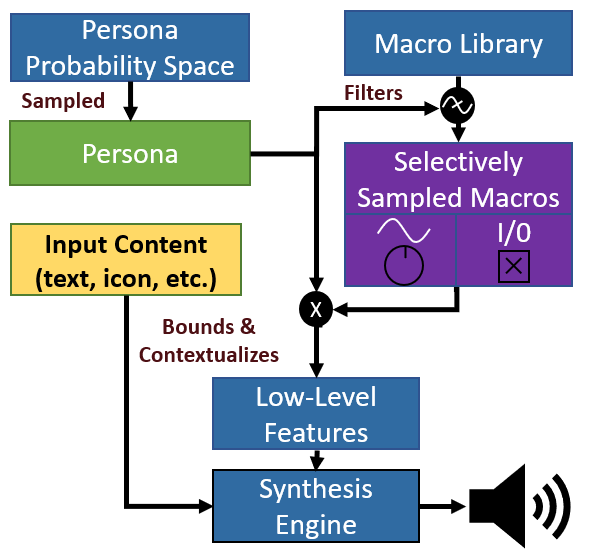}}}
     \caption{A sampled vocal persona parameterizes the representation space of latent variables/low-level features required by a speech synthesis engine. A set of $K$ user-selected expression \textit{macros} allows for modification of this parameterization.}
      \label{fig:flowchart}
\end{figure}

In Figure \ref{fig:flowchart}, we present a visualization of relationship between the \textit{persona probability space} $\mathbf{P}$ and low-level features/latent variables $\mathbf{Z}$ used to synthesis a voice.
We present this framework generally, allowing for the low-level feature space to consist of hand-crafted synthesis parameters or a set of learned latent variables.
We characterize this \textit{persona probability space} $\mathbf{P}$ as a distribution of parameters that describes an $N$-dimensional probability mixture model describing low-level synthesis features $\mathbf{Z} = \{Z_1,Z_2, Z_n,..., Z_N\}$. 
Sampling persona $P_a$ defines the set of $N$ probability density functions (PDF) $f_{a}(z_n|\mathbf{\theta}_{n_a})$ for each synthesis parameter $Z_n$, where $\mathbf{\theta}_n$ are the parameters describing the PDF.  
Sampling a different persona $P_b$ defines another set of $N$ PDFs $f_{b}(x_n|\mathbf{\theta}_{n_b})$ for each feature $Z_n$. 
These distribution spaces could be as overlapped as is perceptually meaningful for the user.
The right column of the flowchart in Figure~\ref{fig:flowchart} shows how perceptually-meaningful expressivity attributes affect the low-level features utilized by a speech synthesis engine.  We propose the concept of a \textit{macro} as a perceptually-informed abstraction that modifies the low-level features such that the modification yields a vocal tone aligned with the intended expressivity.  
For example, a user may want to modify how ``stern" the voice sounds within the bounds of current persona $P_a$. 
A ``stern" control gives the user the ability to modify the ``amount" of sternness in their current voice on a scale from 0 to 100. Given control variable $x \in X \sim \textit{Uniform}[0, 100]$, a function $m_n(x) = w_n y_{n}(x)$ maps $x$ to a corresponding modification value applied to PDF parameters $\theta_n$. 
Here, $w_n$ is a scalar weight corresponding to the involvement of synthesis feature $Z_n$ in the high-level ``stern" macro $M$. 
$y_n(\cdot)$ is a transformation that allows for the weighting of each macro to be configurable and potentially learned or selected by the user. 
Macro $M$ is the set of functions $m_n(\cdot) \forall n \in [1...N]$. Within a persona, a set of $K$ macros $\{M_1,...,M_k,...,M_K\}$ can be created by or presented to the user that allow for modification that is useful or meaningful.  
Within our current proposed design, these macros multiplicatively combine to determine the modification to $\theta_n$.  More explicitly, a user-determined set of $K$ macros can modify the parameters $\mathbf{\theta}_{n_a}$ describing a PDF $f_{a}(z_n|\mathbf{\theta}_{n_a})$ within the current persona mixture model $P_a$ such that:

\begin{equation}
    \theta_{n_a} = (\prod_{k=1}^{K} m_{n_k} (x_k)) \theta_{n_a}, \forall n \in [1...N].
\end{equation}

The ability of users to dictate the level and complexity of interactions with their vocal personas is of paramount importance.
The proposed framework aims to maximize user agency in speech synthesis by allowing for hierarchical user-desired control.


\section{Preliminary Insights and Future Work}\label{sec:discussion}
In addition to a deepened understanding of vocal persona, several design insights were generated over the course of conducting the interview study in parallel with developing our theoretical synthesis framework. 
These design insights include multi-modal communication feedback mechanisms that could be implemented to promote a more embodied experience of vocal communication when using a synthesized voice. 
Insights regarding the disconnect between internal self-perception of one's voice and external perception of a recording highlight the possible need for a separate playback system for users that emulates internal self-perception more closely, possibly including additional feedback elements, such as haptics to promote feelings of embodiment~\cite{HolbrowVocalVoice}. 
Increased environmental awareness, such as listening and responding to changes in room acoustics, background noise level, and other auditory events, could guide context-aware grounding and synthesis, such as pausing and repeating a phrase if a loud, interrupting noise is detected.
Additional exploration of the role of temporal context may also provide valuable insights as long-term contextual variation seem to influence the adoption of certain vocal personas, while short-term variation seems to influence production choices within a persona.

The ability to adjust the voice with greater personalized control over identity-associated features and personality characteristics is a crucial mechanism for enhancing and augmenting user experience, especially within the augmentative and assistive communication (AAC) community~\cite{Pullin2017Designing}. 
We use this conjecture as a main design tenant of the aforementioned persona-based synthesis framework, and are developing this hypothesis further.
Additionally, we are continuing interviews with additional performers, AAC users, conversation designers, and speech scientists to refine our understanding of both natural and synthesized persona.

\small{
\bibliographystyle{ieee}
\bibliography{shared_bib,cam_mendeley}
}
\end{document}